\documentstyle[12pt]{article}
\begin{document}
\setlength{\unitlength}{1mm}
{\hfill   February 1996 }

{\hfill   Alberta Thy 06-96 }

{\hfill    hep-th/9602105 }
 \vspace*{1cm} \\
\begin{center}
{\Large\bf On  One-loop Quantum Corrections to the Thermodynamics of  
Charged 
Black Holes  }
\end{center}
\begin{center}
{\large\bf Valeri P.~Frolov$^{(1,2)}$, Werner Israel$^{(1)}$, Sergey  
N.~Solodukhin$^{(3)}$}
\end{center}
\begin{center}
{\it $^{(1)}$ CIAR Cosmology Program, Theoretical Physics Institute,  
Department of Physics,\\ University of Alberta, Edmonton, Alberta,  
Canada T6G 2J1} \\
\end{center}
\begin{center}
{\it $^{(2)}$ P.N.Lebedev Physics Institute, Leninskii Prospect 53,  
Moscow 117924, Russia}
\end{center}
\begin{center}
{\it $^{(3)}$ Department of Physics, University of Waterloo,  
Waterloo, Ontario, Canada N2L 3G1} \\
%\end{center}
%\begin{center}
{\it and Bogoliubov Laboratory of Theoretical Physics, Joint  
Institute for
Nuclear Research, Head Post Office, P.O.Box 79, Moscow, Russia}
\end{center}
\vspace*{0.5cm}
\begin{abstract}
Quantum corrections are studied for a charged black hole in a
two-dimensional model obtained by spherisymmetric
reduction of the 4D Einstein-Maxwell theory. The classical  
(tree-level)
thermodynamics is re-formulated in the framework of the off-shell  
approach,
considering systems at arbitrary temperature. This implies a conical  
singularity at 
the horizon and modifies the gravitational action by terms defined on
the horizon. A consistent variational procedure for the action   
functional is
formulated. It is shown that  the free energy reaches an extremum on  
the regular
manifold with $T=T_H$. The one-loop contribution to the action in the  
Liouville-Polyakov
form is re-examined. All the boundary terms are taken into account  
and the
dependence on the state of the quantum field is established. The  
modification of 
the Liouville-Polyakov term for a 2D space with a conical defect is  
derived. The backreaction of the
Hawking radiation on  the geometry is studied and the  
quantum-corrected black hole metric is
calculated perturbatively. Within the off-shell approach the one-loop  
thermodynamical
quantities, energy and entropy, are found. They are shown to contain  
a part due to hot gas surrounding
the black hole and a part due to the hole itself. It is noted that  
the contribution of the hot gas
can be eliminated by appropriate choice of the (generally, non-flat)  
reference geometry.
The deviation of the {\it `` entropy - horizon area''}  relation for
the quantum-corrected black hole from the classical law is  
discovered and possible physical
consequences are discussed.
\end{abstract}
\begin{center}
{\it PACS number(s): 04.60.+n, 12.25.+e, 97.60.Lf, 11.10.Gh}
\end{center}
\vskip 1cm
%\noindent $^{ \ast}$ e-mail: frolov@phys.ualberta.ca
%\noindent $^{ \ast\ast\ast}$ e-mail:
%\noindent $^{ \ast\ast\ast}$ e-mail: solod@thsun1.jinr.dubna.su
\newpage
\baselineskip=.8cm
\section{Introduction}
That black holes possess some properties of a thermodynamical system  
characterized by
appropriately defined energy, temperature and entropy 
was first considered as just an analogy
\cite{1} between  black hole physics and the laws of thermodynamics.  
However, the remarkable discovery by
Hawking of radiation from a black hole which  looks
 thermal at large distances \cite{2} strongly supported this analogy  
and forced
physicists  to think of a black hole as a real thermodynamical object  
like a heated
black body. One of the remarkable predictions of the analogy is that  
one can associate entropy  with a hole 
which in the Einstein  theory of gravity is proportional to the area  
of the horizon. Moreover, in  processes involving a hole its entropy  
plays a role on an equal footing with the entropy of conventional   
matter. In particular, only their sum is the quantity which is  
non-decreasing \cite{1,3}.
However, it is a mysterious and intriguing puzzle just what states of  
the hole are counted by the Bekenstein-Hawking entropy. 
As a possible answer one can relate it to states
of quantum fields which are hidden by the horizon and, consequently,  
remain invisible  to an outside observer.
The present status of the problem and numerous attempts  towards its
resolution have been recently reviewed in \cite{4}-\cite{5}.

The role of quantum effects in black hole physics is two-fold.
Semiclassically, a hole can be considered as surrounded by the  
quantum Hawking radiation which becomes thermal (heat bath)
far away the hole. Since this radiation possesses a non-trivial  
stress-energy tensor its backreaction leads to deformation of the  
classical black hole geometry.
On the other hand, the quantum corrections lead to modifications of  
the gravitational effective action. This results in changes to the
formulas for calculating the energy and entropy of the hole.
As an example of such a modification it was recently observed, in two  
\cite{6}, \cite{SS} and in four \cite{F1} dimensions, that
the classical Bekenstein-Hawking expression might be corrected
by terms logarithmically dependent on the mass of a black hole.
The calculations apply the conformal anomaly argument and take a  
fixed classical black hole background. However, the quantum  
deformation of the geometry affects  the black hole parameters, like  
the radius of horizon, introducing some corrections. 
These also turn out to be of the order $\sim \ln M$ and cannot be
neglected. Hence, the backreaction effects necessarily must be
included  when considering  the quantum  thermodynamics of the black  
hole \cite{9}, \cite{SS1}.

Two dimensional physics gives us an arena  (see Refs. \cite{2D},  
\cite{F2}, \cite{M}) where the above-noted
problems can find a precise solution. The 2D non-local
Liouville-Polyakov action \cite{Polyakov} incorporates both the  
Hawking radiation \cite{FV} and its backreaction on the geometry (see  
e.g. Ref. \cite{BD}). Therefore, its inclusion in the gravitational  
action on an equal footing with the classical counterpart gives the  
complete semiclassical description for the black hole.   It is known  
but
not always stressed that the Liouville-Polyakov action contains some  
ambiguity which is eliminated by specification the state of the  
quantum field. In the case of a black hole in  equilibrium with  
thermal radiation this specification
must include the heat bath  at large distances from the black hole.  
As a result, the effective action becomes dependent on the thermal  
state of the quantum field. In principle, this state can be
characterized by a temperature different from the Hawking one. It is  
a remarkable and long-standing fact that such a state can   
effectively be described as a quantum field on a singular instanton  
(i.e. on the Euclidean black hole instanton with conical singularity  
on the horizon).
This probably explains why the Euclidean conical singularity method  
\cite{11}, \cite{Teitelboim}, \cite{F1}, \cite{SS}  gives a sensible  
formulation of  black hole thermodynamics. 
In this method one takes the Gibbons-Hawking \cite{GH} Euclidean  
approach and closes Euclidean time with arbitrary period $\beta$,  
related to the
temperature $T$ of the system as $\beta={1 \over T}$.
Evaluating the free energy $F$ of the system for arbitrary $\beta$,
differentiating $F(\beta )$ with respect to $\beta$
and finally putting $\beta$ equal to the Hawking value  
$\beta=\beta_H$, one obtains the thermodynamical quantities (energy  
and entropy) of the black hole  \cite{SS}, \cite{F1}, \cite{SS1}.

In this paper we use the two-dimensional model to study the one-loop   
quantum effects in the thermodynamics of a charged black hole. We  
start in Section  2 with the 4D Einstein-Maxwell theory with boundary  
terms included appropriately \cite{GH}. Then, considering only  
spherically symmetric metrics, this model reduces to an effectively  
two-dimensional one of the dilaton type. The classical solution  
describes the well-known Reissner-Nordstrom charged black hole.  The  
thermodynamics of the classical black hole is re-formulated in  
Section 3 in the framework of the conical singularity method. We  
especially notice the role of both the terms defined on the external  
boundary and on the conical singularity
in the well-defined variational procedure. The choice of  state of  
the quantum field and the corresponding form of the  
Liouville-Polyakov action is discussed in Section 4. In particular,  
we take care of  the boundary terms and derive the modified  
Liouville-Polyakov action for a space with a conical defect. The  
deformation of the geometry
of a charged black hole  due to
Hawking radiation is calculated perturbatively in Section 5. The  
energy and entropy of the quantum-corrected black hole are calculated  
in Section 6. The deviations from the classical Bekenstein-Hawking  
form are obtained and the possible role of these corrections
is discussed.

\bigskip

\section{Spherically symmetric reduction of 4D Einstein-Maxwell  
theory}
\setcounter{equation}0

Let us consider  4D Einstein gravity coupled with a Maxwell field
described by
the following action (we use the Euclidean signature):
\begin{equation}
W_{cl}=-{1 \over 16\pi G} \int_{M^4}^{}R^{(4)}\sqrt{g}d^4x+
{1 \over 16\pi G}\int_{M^4}^{}F^2_{\mu\nu} \sqrt{g}d^4 x-
{1 \over 8\pi G}\int_{\partial M^4}^{}K^{(4)}\sqrt{h}d^3x ,
\label{2.1}
\end{equation}
where $R^{(4)}$ is  the 4D scalar curvature.
 We have added in (\ref{2.1}) the boundary term according to
\cite{GH}.
 $K^{(4)}$ is the trace of the extrinsic curvature of the
boundary $\partial M^4$. If $n^{\mu}$ is the outward  unit vector  
normal to
$\partial M^4$,
then we have
\begin{equation}
K^{(4)}=\nabla_\mu n^\mu .
\label{2.2}
\end{equation}
The action (\ref{2.1}) is known to be divergent when the boundary  
$\partial M$
goes to infinity.The same presumably happens for the one-loop  
effective action and
requires some subtraction procedure. Generally, one proceeds by  
comparing the divergent quantity with that defined for a specially  
chosen  background. If $g^0_{\mu\nu}$ is the background metric
then we define the subtracted expression as follows \cite{HH}:
\begin{equation}
W_{sub}=W[g_{\mu}]-W[g^0_{\mu\nu}]~~,
\end{equation}
where $W$ includes  both the classical (\ref{2.1}) and one-loop  
gravitational action.
Presumably, in the quantum case we would have
 to subtract the contribution of
the non-flat
reference metric of
 the asymptotic
geometry
(see ref.\cite{SS1} for such a example). Therefore, we shall consider
an arbitrary reference (background)
metric hereafter.

Our first goal is to make the reduction of the general action
(\ref{2.1}) to the special case of spherically symmetric spacetimes.
Spherically symmetric metrics are of
the form
\begin{equation}
ds^2=\gamma_{\alpha\beta}(z)dz^\alpha
dz^\beta+r^2(z)(d\theta^2+\sin^2 \theta d\varphi^2) .
\label{2.3}
\end{equation}
Here $\alpha,\beta,\ldots=0,1$,\ \  $\gamma_{\alpha\beta}(z)$ is  the  
2D
metric on the effective two-dimensional
space $M^2$ covered by coordinates $z^{\alpha}=(\tau, x)$, and   
$r^2(z)$
is the scalar field on
$M^2$. We have for the scalar curvature of the metric (\ref{2.3})
\begin{equation}
R^{(4)}=R^{(2)}+{2 \over r^2} (\nabla r)^2-{2 \over r^2} \Box r^2 +{2
\over r^2} ,
\label{2.4}
\end{equation}
where all the geometrical objects $R^{(2)},~\nabla,~\Box$ are defined
with
respect to 2D metric $\gamma_{\alpha\beta}(z)$.

For the spherical reduction of the action it is sufficient to
consider   boundaries  $\partial M^4$ of the spherically-symmetric
space $M^4$ with metric
(\ref{2.3}) that are a direct product $\partial M^4=\partial M^2
\times
S^2$ where $\partial M^2$ is a boundary of 2D space $M^2$; $S^2$ is a
2D sphere.
A normal vector $n^\mu$ to this boundary has non-zero components only
in the direction tangent to of the space $M^2$,
$n^\mu=(n^\alpha, 0,0)$. Hence, we obtain for the trace of the  
extrinsic curvature of the
boundary (\ref{2.2}):
\begin{eqnarray}
&&K^{(4)}=k+2 n^\alpha{\partial_\alpha r \over r}; \nonumber \\
&&k\equiv \nabla_\alpha n^\alpha\equiv {1\over
\sqrt{\gamma}}\partial_{\alpha}(\sqrt{\gamma}n^{\alpha})=\partial_\alpha
 n^\alpha +{1 \over 2\gamma} \partial_\alpha \gamma
n^\alpha;
\label{2.5}
\end{eqnarray}
where $\gamma=det \gamma_{\alpha\beta}$.  If the metric is  static  
and
spherisymmetric
 it can be  written in
the Schwarzschild form:
\begin{equation}
ds^2=g(x)d\tau^2+g^{-1}(x)dx^2+r^2(x)(d\theta^2+\sin^2 \theta d
\varphi^2).
\label{2.6}
\end{equation}
Then we have $n^\alpha=(0,g^{1/2})$
and hence
$$
K^{(4)}=k+{2 \over r} r' g^{1/2}~~,~~~~k=(g^{1/2})'~~.
$$

In accordance with our assumption about spherical symmetry the
Maxwell field $A_\mu$ is  tangent  to the space $M^2$, i.e. the
only non-zero component of the gauge curvature
is $F_{\tau r} \neq 0$.

Taking into account that  the integration over angles $(\theta,~
\varphi )$ in
(\ref{2.1})  induces the measure
$$
\int_{}^{}\sqrt{g} d\theta d\varphi=4\pi r^2 \sqrt{\gamma}
$$
we finally get that the action (\ref{2.1}) for the spherically
symmetric
metric (\ref{2.3}) reduces to the effective two-dimensional theory
\begin{eqnarray}
W_{cl}=-{1 \over 4G} \int_{M^2}^{}(r^2 R+2(\nabla  
r)^2+2)\sqrt{\gamma}d^2z
+{1 \over 4G}\int_{M^2}^{}r^2 F^2_{\alpha\beta}\sqrt{\gamma} d^2z
-{1 \over 2G}\int_{\partial M^2}^{}r^2 k .
\label{2.7}
\end{eqnarray}
In two dimensions $F_{\alpha\beta}$ has only one component
\begin{equation}
F_{\alpha\beta}=e_{\alpha\beta}F ,
\label{2.8}
\end{equation}
where $e_{\alpha\beta}$ is the antisymmetric Levi-Civita tensor. It
follows
from the equations of motion for the Maxwell field
$$
\nabla_\alpha (r^2 F^{\alpha\beta})=0
$$
that
\begin{equation}
F={Q \over r^2};~~ Q=\mbox{const} ,
\label{2.9}
\end{equation}
where $Q$ is the electric charge.

Inserting (\ref{2.8}), (\ref{2.9}) into the action (\ref{2.7}) we  
find
 that
the whole theory reduces to  some type of  2D dilaton gravity
\begin{eqnarray}
W_{cl}=-{1 \over 4G} \int_{M^2}^{}(r^2 R+2(\nabla  
r)^2+2U(r))\sqrt{\gamma}d^2z
 -{1 \over 2G}\int_{\partial M^2}^{}r^2 k ,
\label{2.10}
\end{eqnarray}
with the field $r^2(z)$ playing the role of the dilaton field. The  
dilaton potential
reads
\begin{equation}
U(r)=1-{Q^2 \over r^2} .
\label{2.11}
\end{equation}
Wick's rotation  to the Euclidean metric is typically accompanied by  
the corresponding
complexification of the charge $Q\rightarrow i Q$ assuming that after  
all calculations we make the continuation back to the real $Q$  
\cite{hawking}. Having this in mind we use the expressions  
(\ref{2.10}) and  (\ref{2.11}) where $Q$ is already real.
Variation of the action (\ref{2.10}) with respect to the dilaton  
$r^2$  
gives the
dilaton equation of motion
\begin{equation}
r R-2\Box r +U'_r=0 ,
\label{2.12}
\end{equation}
while the variation with respect to the metric $\gamma_{\alpha\beta}$  
gives
\begin{equation}
G_{\alpha\beta}\equiv-2r\nabla_\alpha \nabla_\beta  
r+\gamma_{\alpha\beta}(\Box r^2-(\nabla r)^2-U)=0 .
\label{2.13}
\end{equation}
Eq.(\ref{2.13}) implies that the vector $\xi_\alpha=e_\alpha^{
\ \beta}\partial_\beta r$ is a Killing  vector. In the region where  
$(\nabla r)^2\ne 0$ the Killing time $t$  
($\xi^{\alpha}\partial_{\alpha}=\partial_t$) and $r$ can be used as  
coordinates on $M^2$. The equation $G^{\tau}_{\tau}-G^r_r=0$ implies  
that  the metric is of the form
\begin{equation}
ds^2=g(r)d\tau^2+{1 \over g(r)}dr^2 .
\label{2.15}
\end{equation}

The trace of Eq.(\ref{2.13}) is 
\begin{equation}
\Box r^2=2U(r) .
\label{2.14}
\end{equation}
This relation gives
\begin{eqnarray}
g(r)=g_{cl}(r)={1 \over r}\int_{}^{r}U(r')dr' =1-{2M G \over r}+{Q^2  
\over  
r^2}={(r-r_+)(r-r_-)\over r^2} ,
\label{2.16}
\end{eqnarray}   
where $M$ is an integration constant to be identified with the ADM  
mass,
and $r_\pm=MG\pm\sqrt{(MG)^2-Q^2}$ are the radii of the outer and  
inner horizons.

\bigskip

\section{Tree-level black hole thermodynamics}
\setcounter{equation}0

The Euclidean action (\ref{2.10}) is the starting point for the  
formulation
of the classical thermodynamics of the black hole. The standard  
procedure for describing the thermodynamical
properties of a field system  is to go  to the Euclidean space by  a  
Wick's rotation
$t=i \tau$ and to close the $\tau$-direction with period  
$2\pi\beta=T^{-1}$,
where $T$ is  the temperature of the system. The system is assumed to  
be  contained  in  a box of size $L$.  In principle, the field  
configuration does not necessary satisfy
any field equations. The latter arise as a requirement of extremality  
of the free energy functional under appropriately defined boundary  
conditions.

Analogously the thermodynamics  of  black holes can be formulated   
off-shell. We discuss now this formulation in more detail.
 Consider the Euclidean  static metric  of the general type:
\begin{equation}
ds^2=g(x)d\tau^2+{e^{-2\lambda (x) }\over g(x)} dx^2 ,
\label{1}
\end{equation}
written in the fixed coordinate system $(\tau , x)$ where the  
coordinates range between the limits  $0 \leq \tau \leq 2\pi  
\bar{\beta}$; $x_+ \leq x \leq L$. In what follows we assume that an  
external boundary is located at $x=L$, while $x=x_+$ is the location  
of the horizon of the black hole.

The temperature $T$ of the system is fixed at the boundary and  can  
be invariantly defined as
$T^{-1}=\int_{}^{}d\tau g^{1/2}_{00}(x=L)$.
The system is also characterized by the value of the dilaton field  
$r_B$ at the boundary, $r_B=r(x=L)$. The fact that the system  
includes a non-extremal black hole means that at some point  $x=x_+$  
(horizon)  the function $g(x)$ has a simple zero, $g(x_+)=0$. In this  
case the Euler characteristic  of the space described  
by (\ref{1}) is fixed to be $\chi = 1$.
Thus the system is   specified by {\it 1)} fixing  temperature $T$  
and value of the 'radius' $r_B$ on the external boundary, and by   
{\it 2)} fixing black-hole topology.
The statistical ensemble consists of all the functions $(g, \lambda ,  
r)$ satisfying these conditions. For an arbitrary  metric from this  
class
the quantity $\beta_H \equiv  ({2 e^{-\lambda}/g'})_{x=x_+}$ is a  
functional of the metric and it is not fixed by the above conditions.  
In general case such a metric describes  the Euclidean space with  
conical singularity at the point $x=x_+$ (horizon) with angle deficit  
$\delta=(1-{\alpha})2\pi$, where  $\alpha={\bar{\beta }/\beta_H}$.
This implies  that the scalar curvature
has a $\delta$-like contribution coming from the tip of the cone (see  
details in ref.\cite{FS}):
\begin{equation}
R^{(2)}=2({1-\alpha \over \alpha})\delta  
(x-x_+)+\bar{R}^{(2)},~~\alpha={\bar{\beta }\over \beta_H} ,
\label{2}
\end{equation}
where $\bar{R}^{(2)}$ is  the regular part of the curvature. The  
conical singularity vanishes when $\alpha=1$. 
Note that only combination $\alpha={\bar{\beta }/ \beta_H}$
has an invariant meaning while $\beta_H$ and $\bar{\beta}$ are  
coordinate dependent.

In many respects, the approach which we use here  is similar to the
approach developed by York and collaborators \cite{Y}. The essential  
difference however is that in \cite{Y}
only regular metrics are considered.
In our approach  the statistical ensemble  specified by conditions  
{\it 1)},  {\it 2)}
includes both the regular metrics and metrics with conical  
singularities. For a metric of general type (with an arbitrary  
$\alpha$) the classical action (\ref{2.10}) due to (\ref{2}) takes  
the form:
\begin{eqnarray}
W_{cl}&=&-{1 \over 4G} \int_{\bar{M}}^{}(r^2 \bar{R}+2(\nabla  
r)^2+2U(r))\sqrt{\gamma}d^2z
 -\nonumber \\
&&{1 \over 2G}\int_{\partial \bar{M}}^{}r^2 k^{(2)}
-{\pi r^2_+ \over G}(1-\alpha) .
\label{3}
\end{eqnarray}
For the static metric (\ref{1}), action (\ref{3}) is
\begin{equation}
W_{cl}=-{(2\pi\bar{\beta} )\over 4G} \int_{x_+}^{L}( ( r^2)'  
e^\lambda  g'+ 2ge^\lambda (r'_x)^2 +2U e^{-\lambda}) -{\pi r^2_+  
\over G} .
\label{4}
\end{equation}

One can define the free energy $F$, entropy $S$ and energy $E$  
associated with $W_{cl}$ as follows
\begin{equation}
F=(2\pi\beta)^{-1}W_{cl},\hspace{.5cm}S=(\beta\partial_\beta-1)W_{cl}, 
\hspace{.5cm}E={1 \over 2\pi}\partial_\beta W_{cl} ,
\label{6}
\end{equation}
where $2\pi\beta=T^{-1}$ and $\beta=\bar{\beta} g^{1/2}_B$.
Applying these formulas to (\ref{4}) we obtain that the energy $E$ is  
given by
the expression 
\begin{equation}
E=-{1\over 4G g^{1/2}_B}\int_{x_+}^L \left( (r^2)' e^\lambda g'+
2ge^\lambda (r'_x)^2+2Ue^{-\lambda} \right) ,
\label{5}
\end{equation}
and the  entropy  
\begin{equation}
S_{BH}={\pi r^2_+ \over G} 
\label{7}
\end{equation}
takes the standard Bekenstein-Hawking form. In the calculations made  
up to this point  we did not assume that $\alpha=1$, in other words  
the calculations were done off-shell. Now, we fix the temperature  
$T=(2\pi\beta )^{-1}$ and consider the extremum
of the  free energy $F=E-TS$ or equivalently the extremum of the  
action $W_{cl}$.
Remarkably, such an equilibrium configuration automatically satisfies  
the 2-nd law of
black hole thermodynamics:
\begin{equation}
\delta E=T \delta S
\label{8}
\end{equation}
for small variations around the equilibrium state.

It should be noted that only  $T$ and $r_B$ at $x=L$ and
 condition $g(x_+)=0$ at the horizon are assumed to be fixed. The  
functions $g(x),~g'(x),~r(x)$
and the values on the horizon of $r_+=r(x_+),~~g'(x_+)$ ( or $\beta_H  
)$ are variable.
The total variation of the
action $W_{cl}$ is $\delta W_{cl}=\delta_r W_{cl}+\delta_g W_{cl}  
+\delta_\lambda W_{cl}$.
For partial variations we have
\begin{eqnarray}
\delta_r W_{cl}&=&-{2\pi r(x_+) \over G}(1-\alpha )\delta  
r(x_+)\nonumber\\
&&-{(2\pi\bar{\beta}) \over 4G} \int_{x_+}^{L}\delta r (
-2r (e^\lambda  g' )' -4 (g e^\lambda  r')'+2U'_r e^{-\lambda})dx ,
\label{9}\\
\delta_g W_{cl}&=&-{(2\pi \bar{\beta }) \over 4G}
\int_{x_+}^L \delta g ( 
- (e^\lambda  ( r^2)')'+2e^\lambda
r'^2_x )dx ,
\label{10}\\
\delta_\lambda W_{cl}&=&-{(2\pi \bar{\beta }) \over 4G}
\int_{x_+}^L \delta \lambda ( 
e^\lambda  ( r^2)'  g' +2e^\lambda g (r'_x )^2-2Ue^{-\lambda} ) dx .
\label{10'}
\end{eqnarray}
We see that variation of $W_{cl}$ contains terms due to variations of  
the functions $(r,~g,~\lambda )$ inside the region $x_+ \leq x \leq  
L$ that leads  to the equations of motion:
\begin{eqnarray}
-2r (e^\lambda g' )'-4  (g e^\lambda  r')'+2U'_r e^{-\lambda}=0  
,\nonumber \\
-(e^\lambda  ( r^2)' )'+2e^\lambda 
r'^2_x  =0 ,\nonumber \\
e^\lambda  ( r^2)'  g' +2e^\lambda g (r'_x )^2-2Ue^{-\lambda} =0 ,
\label{10''}
\end{eqnarray}
which  of course coincide with equations (\ref{2.12}), (\ref{2.13})  
written
for the metric (\ref{1}).
Variations $\delta g'(x_+)$ and $\delta g' (L)$ on the boundaries  
$(x_+,L)$ 
are cancelled in (\ref{9})-(\ref{10'}).  This happens because of the  
presence of the 'surface' terms in Eq.(\ref{3})  located on the  
external boundary and on the singular point (the cone tip).

 In some sense, the tip $\Sigma$ of the cone can be considered as  
some kind of 
boundary additional to $\partial M$ of the space $M$. It is the  
presence of the additional  term located on $\Sigma$ in the  
gravitational action 
(\ref{3}) that makes  the variational procedure
 on the conical space  well defined. The term connected with the tip  
of the cone compensates variations of the normal
derivatives of the metric at $\Sigma$ in the same manner as the  
standard Gibbons-Hawking
terms does at the external boundary $\partial M$. The variation of  
the action contains also term proportional to the variation of the  
'radius'  $r_+$  of the horizon, $\delta r_+$. The requirement  
$\delta_r W_{cl}=0$ gives the condition: $\alpha=1$. 
This is the expected result. It means that the equilibrium state is  
reached on a regular
manifold without conical singularity (Gibbon-Hawking instanton).

The  equations (\ref{10''})
imply that we may choose $r=x$. The metric function $g(r)$ takes the  
form
(\ref{2.16})
\begin{equation}
g(r)={1 \over r}\int_{r_+}^{r}U(\rho)d\rho .
\label{12}
\end{equation}
In particular, we have
\begin{equation}
g(L)={1 \over L}\int_{r_+}^{L}U(r)dr;~~g'_r(L)=L^{-1}U(L)-L^{-1}g(L)~~. 
\label{13}
\end{equation}
On the other hand, on the horizon we have
\begin{equation}
{2 \over \beta_H} \equiv g'_r(r_+)={U(r_+) \over r_+} .
\label{14}
\end{equation}

The energy functional $E$ (\ref{5}) takes the form
\begin{eqnarray}
E={1 \over 2Gg^{1/2}_B}\int_{x_+}^{L}G^0_0dx+E_{surf}~,  
\hspace{.5cm}E_{surf}=-{1 \over 2G} \left( e^\lambda (r^2)'g^{1/2}  
\right)_{x=L},
\label{14'}
\end{eqnarray}
and modulo  the constraint $G^0_0=0$ it reduces to the surface terms  
only.
Equivalently,
we obtain a coordinate invariant expression for the energy  
(\ref{14'}):
\begin{equation}
E=-{1 \over 2\pi\beta} {1 \over G}  \int_{\partial M}^{}r 
n^\alpha  \partial_\alpha r .
\label{14''}
\end{equation}
The quantity (\ref{14''}) is divergent if $\partial M$ goes to  
infinity. The subtraction procedure described in Section 2 leads to  
the result:
\begin{eqnarray}
E&=&E[g]-E[g_0]={1 \over G}\left({1\over 2\pi\beta_0}\int_{\partial  
M}^{}r n^\alpha_0
\partial_\alpha r-{1\over 2\pi \beta} \int_{\partial M} rn^\alpha  
\partial_\alpha r \right)
\partial_\alpha r \nonumber \\
&=&{1 \over G} \left( r  (g^{1/2}_0-g^{1/2}) \right)_{r=L}~~.
\label{14'''}
\end{eqnarray}
Here we have chosen $r_0=r$ for the reference metric. Note that the  
natural condition to be
imposed on the background is that in the limit $L\rightarrow \infty$  
the background temperature
$T=(2\pi\beta_0)^{-1}$ coincides with the black hole temperature  
measured at infinity. This is satisfied if
$g_0= \lim_{L \rightarrow \infty} g(L)$. For an asymptotically flat  
metric
$$
g(L)=1 -{2MG \over L}+ O({1\over L})
$$
we have $g_0=1$.  Hence for the energy 
\begin{equation}
E={L \over G}[1-g^{1/2}(L)]~~
\label{15}
\end{equation}
we find in the limit $L \rightarrow \infty$ that
\begin{equation}
E=M~~.
\label{16}
\end{equation}
It should be noted that formulating the variational procedure for the  
charged metric
we typically need to augment  quantities fixed at the boundary
by a quantity characterizing the Maxwell sector of the model: charge  
$Q$ or potential
$A_0$ \cite{Y}. The variation with respect to $A_{\mu}$ would give us  
the Maxwell
 equations.
Instead of this we first solved the Maxwell sector exactly and all  
the information
about it was collected in the ``dilaton'' potential $U(r)$, then we   
formulated the variational problem only for the gravitational sector.  
These two ways obviously lead to the same results.

The above consideration is valid for an arbitrary potential
$U(r)$ provided its form is fixed.  For the variations
that change the form of the potential $U(r)$ we obtain from  
(\ref{15})
\begin{equation}
\delta E=\delta M-{ 1 \over 2G}\int_{r_+}^{L}\delta U(r)dr .
\label{19}
\end{equation}
For the special choice of the potential $U(r)$ defined by  
Eq.(\ref{2.11}) we reproduce the known form of the second law for a
charged black hole
\begin{equation}
\delta M=T\delta S +{Q  \over G r_+}\delta Q .
\label{20}
\end{equation}
However, the specific form of the potential $U(r)$ is not essential  
for the above consideration.
It can be shown \cite{Kazakov-Solodukhin} that the quantum  
corrections change the form of the
potential $U(r)$ and results in the deformation of the black hole
metric (\ref{12}). Though our methods can deal with such a  
possibility
as well, we do not consider this here.

Special consideration is needed for an extremal black hole.
In this case we have
\begin{equation}
U(r_+)=0;~~g'(r_+)=0 .
\label{21}
\end{equation}
The geometry of an extremal black hole instanton is very different  
from the
non-extremal one. In the metric
\begin{equation}
ds^2=g(r)d\tau^2+g^{-1}(r)dr^2
\label{22}
\end{equation}
$\tau$ can be closed with arbitrary period $2\pi \beta$ not forming  
any singularity.
The horizon lies now at an infinite distance from any other point of  
the instanton manifold.
Near the horizon the extremal instanton resembles a constant  
curvature
space with metric
\begin{equation}
ds^2={r_+^2 \over z^2}(d\tau^2+dz^2) ,
\label{23}
\end{equation}
where $z\rightarrow -\infty$ if $r\rightarrow r_+$. The extremal  
black hole
instanton can be considered as conformally related  to a flat  
cylindrical space.

These features of the extremal geometry are crucial for the
formulation of the thermodynamics of the extremal hole \cite{EX}.  
Since there is no conical
singularity on the horizon we do not have the additional term in the  
action and
it reads
\begin{equation}
W=2\pi\beta E ,
\label{24}
\end{equation}
where the energy $E$ takes the form (\ref{15}). We obtain from  
(\ref{24})
for the free energy of the system $F=E$, and hence the entropy of the  
extremal hole
is formally zero:
\begin{equation}
S_{ext}=0
\label{25}
\end{equation}
Moreover, since the free energy does not depend on the temperature  
$\beta^{-1}$,
the requirement of extremality of the free energy under $\beta$ fixed
does not give a relation between parameters of the hole geometry  
($r_+$) and
$\beta$ as we found  for non-extremal case. This can be interpreted  
as implying that the extremal
black hole can be in equilibrium at arbitrary temperature \cite{EX}.
However the physical meaning of this formal result is not clear. In  
particular,  quantum effects may change this conclusion. We are going  
to consider this in a separate publication.

\bigskip

\section{Liouville-Polyakov action and choice of the thermal state of  
the quantum field}
\setcounter{equation}0

In order to include  one-loop quantum effects  in the analysis,  
consider a
two-dimensional quantum conformal massless scalar field. This  
produces the following contribution
to the partition function:
\begin{eqnarray}
Z=e^{-\Gamma}~~~, \hspace{.5cm}\Gamma={1 \over 2} \ln det \Box~~~,
\label{3.1}
\end{eqnarray}
where $\Box =\nabla_\mu \nabla^\mu$ is the two-dimensional Laplacian.
The calculation of the effective action $\Gamma$ is usually made by
 integrating  the conformal anomaly. The result is well-known  
\cite{Polyakov}:
\begin{equation}
\Gamma_{PL}[g]={1 \over 96\pi}\int_{}^{}R \Box^{-1} R ~~.
\label{3.2}
\end{equation}
However, if we wish to work with  (\ref{3.2}) we are confronted with  
at least two problems.
First, the action (\ref{3.2}) does not transform properly under a  
constant (global)
conformal transformation, $g_{\mu\nu} \rightarrow \Lambda  
g_{\mu\nu}$.
(This was noted by Dowker \cite{D}.) Second, when applying  
Eq.(\ref{3.2})
to a  flat space (where $R=0$), we get the the corresponding mean  
value of the stress-energy tensor $\langle T^{\mu\nu} \rangle$ obtained by  
the variation of Eq.(\ref{3.2}) vanishes. This is certainly valid for  
the vacuum state, but not for other possible states. In particular,   
it is not clear how Eq.(\ref{3.2}) can reproduce  
the effective action for a thermal radiation.
 So, writing the effective action in the form (\ref{3.2}) one loses  
the information
on the concrete choice of the state of the quantum field. We  
demonstrate that the  
information about the state of a quantum field is directly connected  
with the boundary terms which are to be added to Eq.(\ref{3.2}).
Therefore, we begin our consideration of one-loop
quantum effects with a more careful treatment of the  
Liouville-Polyakov action,
taking into account all the boundary terms.

It should be emphasized that the integration of the conformal anomaly  
which is used to derive Eq.(\ref{3.2}) does not give the absolute  
value of the effective action $\Gamma[g]$,
but rather the difference between the effective actions for two  
conformally related
($g_{\mu\nu}=e^{2\sigma} \hat{g}_{\mu\nu}$) manifolds \cite{Alvarez}:
\begin{eqnarray}
\Gamma [g]= \Gamma [\hat{g}]-{1 \over 24\pi} \left( \int_{M}^{}
(\hat{\nabla}\sigma)^2
+\int_{M}^{}\hat{R} \sigma+2\int_{\partial M}^{}d\hat{s}
\hat{k}\sigma
\right)-{1 \over 8\pi} \int_{\partial M}^{}
d\hat{s}\hat{n}^\mu\partial_\mu
\sigma .
\label{3.3}
\end{eqnarray}
Here $\hat{n}^\mu$ is the outward vector normal to the boundary  
$\partial M$, and
$\hat{k}=\nabla_\mu \hat{n}^\mu$ is the trace of the second  
fundamental form of the
boundary.

One can write $\Gamma [g]$ in terms of quantities defined only with  
respect to
metric $g_{\mu\nu}$ if we introduce an additional field $\psi$  
defined as a solution of
the equation
\begin{equation}
\Box \psi =R .
\label{3.4}
\end{equation}
For conformally related  metrics  
$g_{\mu\nu}=e^{2\sigma}\hat{g}_{\mu\nu}$
the respective quantities are related as:
\begin{eqnarray}
R=e^{-2\sigma}(\hat{R}-2 \stackrel{\wedge}{\Box} \sigma  
),\hspace{.5cm}\psi=\hat{\psi}-2\sigma ,\nonumber \\
k=e^{-\sigma} ( \hat{k}+\hat{n}^\mu \partial_\mu \sigma )  
,\hspace{.5cm}n^\mu=e^{-\sigma} \hat{n}^\mu .
\label{3.5}
\end{eqnarray}
Using these relations, one can show that the effective action $\Gamma  
[g]$ of (\ref{3.1}),
conformally transforming according to (\ref{3.3}), takes the form:
\begin{eqnarray}
\Gamma [g]={1 \over 48 \pi} \int_{M}^{} ({1 \over 2} (\nabla \psi  
)^2+
\psi R )+{1 \over 24\pi }\int_{\partial M}^{}k \psi ds  +\Gamma_0~~,
\label{3.6}
\end{eqnarray}
where all the quantities are defined with respect to $g_{\mu\nu}$ and
the "integration constant" $\Gamma_0$ is a conformally invariant  
functional.

\bigskip

Let us now consider  the conformal massless  field $\varphi$  in a  
thermal
state with temperature $T$ in a space-time with horizon. The relevant  
static Euclidean metric
reads
\begin{equation}
ds^2=g(x)d\tau^2+{1 \over g(x)}dx^2 ,
\label{3.16}
\end{equation}
or
\begin{equation}
ds^2=g(\rho)d\tau^2+d\rho^2 ,
\label{3.17}
\end{equation}
where $\tau$ lies in the range $0\leq \tau \leq 2\pi\bar{\beta}$ and  
$0\leq \rho \leq L_\rho$.
Assume that $g(x)$ has a zero of  first order at the point $x=r_+$.  
This is the Killing
horizon. Near the horizon we have $g(\rho)={\rho^2 / \beta^2_H}$,  
where
$\beta_H=2/g'_x(r_+)$. For $\bar{\beta}=\beta_H$ Eq.(\ref{3.17})   
describes
a regular black hole instanton. If $\bar{\beta} \neq \beta_H$
 the metric has a conical singularity at $\rho=0$ with angle deficit  
$\delta=2\pi(1-\alpha),~~
 \alpha={\bar{\beta} /\beta_H}$. The metric (\ref{3.17}) can be  
written in  
the conformal
 form:
\begin{eqnarray}
ds^2&=&e^{2\sigma}ds_0^2,\hspace{.5cm}ds_0^2=(dz^2+\alpha^2 z^2  
d\tilde{\tau}^2)~~, \nonumber \\
e^{2\sigma}&=&\beta^2_H {g \over z^2}~~, 
z=z_0\exp\left[{{1 \over \beta_H} \int_{L_\rho}^{\rho}{d\rho \over  
\sqrt{g}}}\right] ,
\label{3.18}
\end{eqnarray}
where $\alpha={\bar{\beta} / \beta_H}$, $\tau =\bar{\beta}  
\tilde{\tau}$, and $0 \leq \tilde{\tau}
 \leq 2\pi$, $0 \leq z \leq z_0$. Note that near the horizon
$z \approx \rho$ and hence the conformal factor is regular on the  
horizon.

For $\bar{\beta}=\beta_H$ Eq.(\ref{3.18}) conformally relates the  
metric of the black hole instanton with the metric on the
flat disk $\cal D$ of radius $z_0$. For conformally related metrics,
$g_{\mu\nu}=e^{2\sigma}\hat{g}_{\mu\nu}$, the stress-energy tensors  
are related
as follows:
\begin{equation}
T_{\mu\nu}[g]=T_{\mu\nu}[\hat{g}] +{1 \over 48\pi} \left(  
-4\hat{\nabla}_\mu
\hat{\nabla}_\nu
\sigma +4\partial_\mu \sigma \partial_\nu \sigma +g_{\mu\nu}  
(4\hat{\Box}
 \sigma-2(\hat{\nabla} \sigma)^2)
\right) .
\label{3.19}
\end{equation}
Thus, for (\ref{3.18}) we have
\begin{equation}
T_{\tau\tau}=
T_{\tau\tau}^{(0)}+{1 \over 48\pi} ({2 \over \beta^2_H}+2g''_\rho-
{3 \over 2} {(g'_\rho)^2 \over g} ) ,
\label{3.20}
\end{equation}
where $T^{(0)}_{\tau\tau}$ is energy density of the quantum field on  
the flat disk
$\cal D$. At infinity $\rho=\infty$\  \  $g=1$, so we have
\begin{equation}
T_{\tau\tau}=T^{(0)}_{\tau\tau}+{1 \over 24\pi\beta^2_H} .
\label{3.21}
\end{equation}
Assume that the quantum field on $\cal D$ is in the state for which
$T^{(0)}_{\tau\tau}=0$. We call this state 'vacuum on the disk'.  
Physically this state is just the usual Minkowski (or Hartle-Hawking)  
vacuum state in the Rindler space.

For this choice we find that the quantum field on the black hole  
instanton is in
the state of the Hartle-Hawking vacuum with  Hawking temperature  
$T_H={1 /2\pi\beta_H}$ since (\ref{3.21})
coincides with the energy density  of a thermal bath
 with temperature $T_H$. Hence,  starting with the 'vacuum on the  
disk' state on the flat
 disk and making the regular conformal transformation (\ref{3.18}),  
we obtain the
quantum field in the state with Hawking temperature on the regular  
black hole instanton.
If we start with the state at finite temperature $T_0=(2\pi\beta_0  
)^{-1}$ on the
disk $\cal D$ we obtain   the state with the temperature
$T=(2\pi\beta)^{-1}= (2\pi)^{-1} [\beta^{-2}_0+\beta^{-2}_H]^{-1/2}$  
on the black hole instanton, which differs from  the Hartle-Hawking  
state.

After these general remarks consider now a singular black hole  
instanton $M^\alpha$ with $0 \leq \tau \leq 2\pi
\bar{\beta}$
~~~$(\bar{\beta} \neq \beta_H)$. Then Eq.(\ref{3.18}) conformally  
relates it to the flat
cone ${\cal C}_\alpha$\ \  ($\alpha={\bar{\beta}/\beta_H}$), and  
$z_0$ is the proper length  of the cone's generator. The conformal  
factor $\sigma$
is an everywhere regular function, and we find that the stress-energy
tensors $T_{\mu\nu}$
on the two spaces
are related by the same expression (\ref{3.19}), (\ref{3.20}) where  
now $T^{(0)}_{\tau\tau}$
is the energy density on the flat cone ${\cal C}_\alpha$ (  
\cite{FZ}, see also \cite{W}):
\begin{equation}
T^{\tau~(0)}_\tau={1 \over 24\pi}{1 \over z^2}({1-\alpha^2 \over  
\alpha^2}),~~\alpha={\bar{\beta} \over \beta_H} .
\label{3.22}
\end{equation}
At infinity, the energy density
$$
T_{\tau\tau} \rightarrow{1 \over 24\pi{\bar{\beta}}^2} 
$$
takes the thermal form with temperature  
$T_{\infty}={2\pi\bar{\beta}}^{-1}$. Hence we may
 conclude that the thermal state of the quantum field with $T\neq  
T_H$
in the gravitational field of a black hole  can be effectively  
described as a quantum field on a singular  instanton (i.e. on the  
instanton with a conical singularity on the horizon).

One can calculate $T_{\mu\nu}$  directly in terms of the metric on  
the
black hole instanton $M^\alpha$ with a conical singularity  
($\bar{\beta} \neq \beta_H$)
(see \cite{SS}).
Eq.(\ref{3.4}) for the metric (\ref{3.17}) has the following  
solution:
\begin{eqnarray}
\psi=- \ln g+b \int_{}^{x}{dx \over g}+C =-\ln g+b  
\int_{}^{\rho}{d\rho \over \sqrt{g}}+C .
\label{3.23}
\end{eqnarray}
In order to fix constant $b$ in (\ref{3.23}) consider
the renormalized stress-energy tensor  which is expressed via $\psi$  
as follows
\cite{Reuter}: 
\begin{equation}
T_{\mu\nu}={1 \over 48\pi} \left(2 \nabla_\mu\nabla_\nu  
\psi-\partial_\mu\psi
\partial_\nu \psi+g_{\mu\nu}(-2R+{1 \over 2}(\nabla \psi)^2)  
\right)~~. 
\label{3.24}
\end{equation}
The conformal transformation of (\ref{3.24}) is given by  
(\ref{3.19}).
Inserting $\psi$ (\ref{3.23}) into (\ref{3.24}) we obtain
\begin{equation}
T_{\tau\tau}={1 \over 48\pi}(2g''_\rho-{3 \over 2}{(g'_\rho)^2 \over  
g}+{b^2 \over 2}) .
\label{3.25}
\end{equation}
In order to have at infinity  thermal behavior with  
$T=(2\pi\bar{\beta})^{-1}$
we must fix the constant $b={2 \over \bar{\beta}}$ in (\ref{3.23}).

This identification automatically gives us that 
in the limit
$\rho \rightarrow 0$  the function $\psi$  (\ref{3.23})
\begin{equation}
\psi\rightarrow\psi_c=-2(1-{\beta_H \over \bar{\beta}}) \ln \rho
\label{3.26}
\end{equation}
 coincides with the  solution of the cone equation
\begin{eqnarray}
\Box_c \psi_c=R_c~~, \hspace{.5cm}R_c=2({1-\alpha \over \alpha})  
\delta (\rho) ,
\label{3.27}
\end{eqnarray}
where $\Box_c$ is the Laplacian on the flat cone ${\cal C}_\alpha$.
Thus, the stress-energy tensor $T_{\mu\nu}$ for the state with  
temperature
$T=(2\pi\bar{\beta})^{-1}$ at infinity coincides with the  
$T_{\mu\nu}$ of a quantum field on the black hole
instanton (\ref{3.18}) with conical singularity on the horizon  
($\bar{\beta} \neq \beta_H$).

In order to fix the constant $C$ in (\ref{3.23}), which in fact can  
depend on the characteristics of the system, consider the conformal  
transformation determined by $\sigma (x)$ (\ref{3.18}):
\begin{equation}
2\sigma (x)= \ln g(x)+{2 \over \beta_H} \int_x^L {dx \over g(x)} +2  
\ln {\beta_H \over z_0}
\label{I}
\end{equation}
which relates (see (\ref{3.18})) our singular black hole instanton  
with a flat cone $C_\alpha$
with radius $z_0$. Then we have that the functions $\psi (x)$ on  
these spaces are related as follows:
\begin{equation} 
\psi_{M^\alpha} (x)=\psi_{C_\alpha}(z)-2\sigma (x)~~,
\label{II}
\end{equation}
where $z(x)$ is given by (\ref{3.18}). On the other hand, for each  
functions
$\psi_{M^\alpha}$ and $\psi_{C_\alpha}$ we have the representation  
(\ref{3.23}):
\begin{eqnarray}
\psi_{M^\alpha}(x)&=&-\ln g(x)-{2\over \bar{\beta}} \int_x^L {dx  
\over g(x)} +C ,\nonumber \\
\psi_{C_\alpha}(z)&=&-2(1-{1\over \alpha}) \ln {z \over z_0}  
+C(\alpha , z_0) .
\label{III}
\end{eqnarray}
Here $C(\alpha , z_0)$ is function of only $\alpha$ and $z_0$.  

Plugging (\ref{I}), 
(\ref{III}) into (\ref{II}) we find for the constant $C=-2 \ln  
{\beta_H \over z_0}
+C(\alpha, z_0)$. Really, there is no dependence of $C$ on $z_0$  
since under rescaling $z_0 \rightarrow e^\gamma z_0$ we have  
$C(\alpha , z_0) \rightarrow C(\alpha , z_0)-2\ln \gamma$.
Thus, finally we have
\begin{equation}
\psi_{M^\alpha}(x)=-\ln g(x) -{2\over \bar{\beta}} \int_x^L {dx \over  
g(x)} -2 \ln {\beta_H \over z_0}+ C(\alpha , z_0) .
\label{IIII}
\end{equation}

\bigskip

In order to write down the Polyakov-Liouville action for this case it  
should be noted that in the presence of the conical singularity
the conformal transformation of the effective action (\ref{3.1}) must  
be modified.  If two conical spaces $M^\alpha$ and
$\hat{M}^\alpha$ with the angle
deficit $\delta=2\pi(1-\alpha)$ and a tip $\Sigma$ are related by a  
regular conformal transformation $g_{\mu\nu}=
e^{2\sigma}\hat{g}_{\mu\nu}$ then the corresponding effective actions   
are related
as follows \cite{D1}:
\begin{eqnarray}
\Gamma [g]&=& \Gamma [\hat{g}]-{1 \over 24\pi} \left(  
\int_{\hat{M}^\alpha}^{}
(\hat{\nabla}\sigma)^2
+\int_{\hat{M}^\alpha}^{}\hat{R} \sigma+2\int_{\partial  
\hat{M}^\alpha}^{}d\hat{s}
\hat{k}\sigma
\right) \nonumber \\
&-&{1 \over 8\pi} \int_{\partial \hat{M}^\alpha}^{}
d\hat{s}\hat{n}^\mu\partial_\mu
\sigma -{1 \over 12}{(1-\alpha^2) \over \alpha}\sigma_h~~,
\label{3.28}
\end{eqnarray}
where $\sigma_h$ is the value at the tip $\Sigma$ of the cone.

Taking into account the transformation law (\ref{3.5}) of $\psi$,   
the effective action
on $M^\alpha$, transforming according to (\ref{3.28}), can be written  
in the form:
\begin{eqnarray}
\Gamma [M^\alpha]&=&{1 \over 48 \pi} \int_{M^\alpha}^{} ({1 \over 2}  
(\nabla \psi )^2+
\psi \bar{R} ) +{1 \over 24}{(1-\alpha^2) \over \alpha}\psi_h  
\nonumber \\
&+&{1 \over 24\pi }\int_{\partial M^\alpha}^{}k \psi ds +\Gamma_0~~.
\label{3.29}
\end{eqnarray}
Here $\bar{R}$ is the regular part of the scalar curvature, and 
$\psi (x)$ in (\ref{3.29}) is the solution of the equation
$\Box \psi=R \equiv 2{(1-\alpha) \over \alpha  
}\delta_\Sigma+\bar{R}$. 
For a static metric (\ref{3.16}) $\psi$  takes the form (\ref{IIII}).  

We denote by
$\psi_h=\psi (\Sigma )$  the value of $\psi$ 
on the horizon (tip of the cone) and  by $\Gamma_0$  a  conformally  
invariant
functional.

It is worthwhile to note that the expression (\ref{3.29}) can be  
rewritten
in two equivalent forms. The first one
\begin{eqnarray}
\Gamma [M^\alpha]={1 \over 48 \pi} \int_{M^\alpha}^{} ({1 \over 2}  
(\nabla \psi )^2+
\psi R ) +{(1-\alpha )^2 \over 24\alpha}\psi_h +{1 \over 24\pi  
}\int_{\partial M^\alpha}^{}k \psi ds +\Gamma_0~~,
\label{3.29''}
\end{eqnarray}
involves quantities defined on the full conical space $M^\alpha$:
$R \equiv 2 (\alpha^{-1}-1)\delta_\Sigma+\bar{R}$, $\Box \psi =R$.

Another way to present the effective action on the conical space
$M^\alpha$ is to write it by using  quantities defined only on the  
regular
part $M^\alpha \setminus \Sigma$:
\begin{eqnarray}
\Gamma [M^\alpha]&=&{1 \over 48 \pi} \int_{M^\alpha \setminus  
\Sigma}^{} ({1 \over 2} (\nabla  \bar{\psi })^2+
\bar{\psi} \bar{R} ) +{1 \over 12}(1-\alpha)\bar{\psi}_h \nonumber \\
&+&{1 \over 24\pi }\int_{\partial M^\alpha}^{}\bar{k} \bar{\psi} ds
+O((1-\alpha )^2)~~,
\label{3.29'''}
\end{eqnarray}
where $\bar{\psi}=\psi_{\alpha=1}, ~~\Box \bar{\psi}=\bar{R}$. The  
effective action in this form was written in \cite{SS1}.

\bigskip

\section{Quantum-corrected black hole geometry}
\setcounter{equation}0

In the semiclassical approximation (when the metric is not quantized)  
the one-loop
quantum effects are taken into account by adding to the classical  
gravitational action
the quantum counterpart obtained by integrating out the matter  
fields:
\begin{equation}
W=W_{cl}+\Gamma .
\label{5.1}
\end{equation}
Following our spherically symmetric considerations we take the  
classical part
$W_{cl}$ to have the form (\ref{2.10}) (with correct subtraction of  
the
contribution due to the reference metric as has been explained in  
Section 3)
while the one-loop contribution
$\Gamma$ is the Polyakov-Liouville action (\ref{3.29}).
Of course, in a self-consistent treatment  the quantum effective  
action $\Gamma$ must be
obtained by the same spherically symmetric reduction of the 4D matter  
fields
as  has been done for the gravitational part $W_{cl}$. However, the  
effective action
becomes a rather complicated quantity which makes  the  analysis  
difficult.
Therefore, we consider here the simplest case when the effective  2D  
matter is
conformal and $\Gamma$ is the non-local Polyakov-Liouville  
functional.

We begin our consideration of one-loop quantum effects  by the  
studying
the corrections to the classical geometry of the black hole induced  
by
quantum  corrections to  the action (\ref{5.1}). Variation of  
(\ref{5.1})
with respect to the metric gives the equations:
\begin{eqnarray}
G_{\alpha\beta}&=&-T_{\alpha\beta}~~, \label{5.2} \\
T_{\alpha\beta}&=&{G \over 24\pi} (2 \nabla_\alpha \nabla_\beta \psi
-\partial_\alpha \psi \partial_\beta \psi
-\gamma_{\alpha\beta}(2R-{1 \over 2}(\nabla \psi )^2)~~;
\label{5.2a}
\end{eqnarray}
where $G_{\alpha\beta}$ is given by Eq.(\ref{2.13}). The  variation
with respect to the dilaton field $r^2(x)$ gives the same equation as  
in the
classical case:
\begin{equation}
2R-2\Box r +U'_r=0 .
\label{5.3}
\end{equation}
An important consequence of  Eqs.(\ref{5.2}) and  (\ref{5.3})
is that the space-time singularity now is placed at finite radius
(value of the dilaton) $r^2=r^2_{cr} \equiv {G \over 12\pi}$.
This typically happens in two-dimensional models
of gravity, as has been previously observed in the string context
\cite{X} and for the theory under consideration in \cite{Lowe},  
\cite{Kazakov-Solodukhin}.
For this value of the dilaton the kinetic term in (\ref{5.1}) becomes  
degenerate.
On the other hand, taking the trace of (\ref{5.2}) we have
\begin{equation}
\Box r^2-2U(r)={G \over 12\pi}R .
\label{5.4}
\end{equation}
Combining this relation with Eq.(\ref{5.3}) we obtain for the  
curvature
\begin{equation}
R={2U-rU'-2(\nabla r)^2 \over r^2-r^2_{cr}} ,
\label{5.5}
\end{equation}
which implies a singularity at $r=r_{cr}$.
We do not investigate here the behavior of the solution of  
Eqs.(\ref{5.2}) and (\ref{5.3})
near this point. Instead, we assume that the outer horizon lies at
$r_+ >> r_{cr}$. Then, in the region $r \geq r_+$ we may solve  
Eqs.(\ref{5.2}) and  (\ref{5.3})
perturbatively (with respect to ${r_{cr}/r_+}$) considering  
$T_{\alpha\beta}$ in the r.h.s. of Eq.(\ref{5.2}) as a small  
perturbation.
This  gives  the correction to the black hole geometry to  first
order in the Planck constant~$\hbar$.

As earlier we consider a static solution. We define  functions $f$  
and $M$ as 
\begin{equation}
f=(\nabla r)^2,\hspace{1cm}M={1\over 2}r(1-(\nabla r)^2)+{Q^2 \over  
r} ,
\label{5.6}
\end{equation}
and choose $r$ as one of the coordinates, while the Killing time $t$  
as the other coordinate. For this choice of the coordinates we get
\begin{eqnarray}
ds^2&=&f(r)e^{2\Phi (r)}dt^2+{1 \over f(r)} dr^2~~, \label{5.9} \\
f(r)&=&1-{2M(r) \over r}+{Q^2 \over r^2} .
\label{5.9a}
\end{eqnarray}
The equation (\ref{5.2}) takes the form
\begin{eqnarray}
2r \nabla_\alpha \nabla_\beta r=\gamma_{\alpha\beta} {2M \over r}
-\gamma_{\alpha\beta}T+T_{\alpha\beta}  
,\hspace{.5cm}T=\gamma_{\alpha\beta}T^{\alpha\beta} .
\label{5.7}
\end{eqnarray}

Differentiating Eq.(\ref{5.6}) and using Eq.(\ref{5.7}) we obtain 
\begin{equation}
2\partial_\alpha M=\partial_\beta r (\delta_\alpha^\beta  
T-T_\alpha^\beta) .
\label{5.8}
\end{equation}
This equation is identically satisfied for the value of index  
$\alpha=0$, while for $\alpha=1$ it gives
\begin{equation}
\partial_r M={1\over 2}T^t_t .
\label{5.10}
\end{equation}
By taking the trace of Eq.(\ref{5.7}) we obtain the equation
for the function $\Phi (r)$:
\begin{equation}
\partial_r \Phi={1 \over 2r f}(T^t_t-T^r_r) .
\label{5.11}
\end{equation}
We consider the r.h.s. of equations (\ref{5.10}) and (\ref{5.11}) as  
a perturbation.
Then, solving these equations perturbatively, we must take
their right-hand sides  on the classical background.
At the classical level we have $\Phi (r)=0$ and $M=\mbox{const}$, and  
for the static
metric (\ref{5.9}) the stress-energy tensor $T_{\alpha\beta}$   
(\ref{5.2})
reads 
\begin{eqnarray}
T^t_{t}&=&\kappa \left( +2f''-{1 \over 2f}(f'^2-{4 \over \beta_H^2})  
\right) ,\nonumber \\
T^r_{r}&=&\kappa \left({1 \over 2f}(f'^2-{4 \over \beta^2_H}) \right)~~. 
\label{5.12}
\end{eqnarray}
Here $\kappa={G /24\pi}$, $\beta_H={2/f'(r_+)}$.  We must put the
classical metric (\ref{2.16}) with  
$f=g_{cl}(r)=r^{-2}(r-r_+)(r-r_-)$,
$r_{\pm}=MG\pm\sqrt{(MG)^2-Q^2}$
 into Eq.(\ref{5.12}). 

It should be noted that $T_{\alpha\beta}$ given by Eq.(\ref{5.12}) is  
divergent
at the inner horizon $r=r_-$. 
This is the well-known divergence \cite{Div} which makes
the perturbation scheme non-applicable near $r=r_-$.
To derive the conditions of applicability of the perturbation scheme  
consider $T^\alpha_\beta$ first at the outer horizon $r=r_+$ and then  
take $r_-\sim r_+$. Then we observe that both $T^t_t$ and $T^r_r$  
defined by Eq.(\ref{5.12}) remain finite in this limit, while the  
combination $f^{-1}(T^t_t-T^r_r)$ appearing in (\ref{5.11}) diverges  
as $\kappa[(r_+-r_-)r_+]^{-1}$.
The perturbation analysis works if the parameters $r_+,r_-$ are such  
that this dangerous term is
eliminated by the condition $\kappa[(r_+-r_-)r_+]^{-1}<<1$ which  
implies that
$\kappa[r_+]^{-2}<<1-{r_-/r_+}$. Thus, taking  $r_+$ to be  large  
enough
we always can come arbitrary close to  extremality $r_-\sim r_+$.  
This important circumstance
allows us apply our consideration to  charged black holes with $Q  
\sim M$ that guarantees stability of the thermodynamical ensemble for  
an arbitrary large 'radius' $r_B$ of the external boundary.

Eqs.(\ref{5.10}), (\ref{5.11}) are easily integrated. Denote
\begin{eqnarray}
m(r)=2\kappa^{-1}(M-M(r)).
\label{5.13}
\end{eqnarray}
then the integration of Eq.(\ref{5.10}) gives
\begin{eqnarray}
m(r)&=&-{1\over \kappa}\int_{}^{r} T^t_t (r)dr=C(r)+A\ln  
{(r-r_-)\over l}+B\ln{r\over l}~~; \nonumber \\
C(r)&=&-{2 \over \beta^2_H}r -{(r_+-r_-)^2 \over 2r_+r_-r}
-{2(r_++r_-)\over r^2}+{10 r_+r_- \over 3r^3}~~,            \nonumber  
\\
A&=&-{(r_+-r_-)^2(r_++r_-)(r_+^2+r_-^2)\over 2r_+^4r_-^2}~~,  
\nonumber \\
B&=&{(r_+-r_-)^2(r_++r_-)\over 2r^2_+r^2_-} .
\label{5.14}
\end{eqnarray}
As earlier $r_\pm$  denotes  the 'radius'  of the classical inner and  
outer horizons.  The following useful identity between constants $A$  
and $B$ is worth noting: $A+B=-{4MG  \beta^{-2}_H}$.
In Eq.(\ref{5.14}) we have introduced a distance $l$ in order to have   
dimensionless
quantities under the logarithms. The final results for the energy and  
entropy
calculated in  Section 6 do not depend on this parameter. It seems   
natural
to assume $l$ to be of  order of the Planckian length $l\sim r_{cr}$.  
However,
this point is not essential for our further considerations.

Similarly the integration of the Eq.(\ref{5.11}) 
\begin{equation}
\Phi (r)={1\over 2}\int_{r}^{L}{1\over rf}(T^r_r-T^t_t)dr .
\label{5.15}
\end{equation}
for $f=g_{cl}(r)$
with the imposed  condition $\Phi (L)=0$ gives
\begin{eqnarray}
\Phi (r)&=&{1\over 2}\kappa \left( F(L)-F(r) \right) ,\nonumber \\
F(r)&=&-{(r^4_+-r^4_-)\over r_+^4 r_-(r-r_-)}+{4\over  
r^2}+{4(r_++r_-)\over
r_+r_- r}+D\ln [(r-r_-)/l]+E\ln (r/l) ,\nonumber \\
D&=&{1 \over r_+^4r_-^2}(3r_+^4+2r^3_+r_-  
+2r^2_+r_-^2+2r_+r_-^3-r_-^4) ,
\nonumber \\
E&=&{1\over r_+^2r_-^2}(-3r_+^2-2r_+r_--3r_-^2) .
\label{5.16}
\end{eqnarray}

Consider now the special case of the {\it uncharged black hole  
($Q=0$)}. The classical metric function  is $g_{cl}(r)=1-{r_+/r}$,  
$r_+=2MG$,
$\beta_H=2r_+$. For the quantum-corrected metric  we get
\begin{eqnarray}
f(r)&=&1-{2MG \over r}+{\kappa m(r)\over r}~~,
\label{5.17}\\
m(r)&=&-{7r_+ \over 4r^2}+{1\over 2r}-{2r \over \beta^2_H}-{1\over  
2r_+}\ln {r\over l} .
\label{5.18}
\end{eqnarray}
and
\begin{eqnarray}
\Phi (r)&=&{1\over 2}\kappa \left( F(L)-F(r) \right) ,\nonumber \\
F(r)&=&{3\over 2r^2}+{2\over r_+r}-{1\over r^2_+}\ln{r\over l}.
\label{5.19}
\end{eqnarray}
For a large size $L$ of the box  we have
\begin{equation}
\exp{(2\Phi (r))}=({r\over L})^{\kappa /r^2_+} \exp{[-\kappa (
{3\over 2r^2}+{2\over r_+r} )]} .
\label{5.20}
\end{equation}

\bigskip

One of the important characteristics of a black hole is the radius of  
its
horizon. In our model its role is played by the value $\bar{r}_+$ of  
the  
dilaton field on the horizon.  For the  quantum-corrected solution  
(\ref{5.13}) it is shifted
with respect to the classical value $r_+$. To see this, take the  
condition  
$f(\bar{r}_+)=0$\
which is solved as follows:  
$\bar{r}_+=M(\bar{r}_+)G+\sqrt{(M(\bar{r}_+)G)^2-Q^2}$.  

Expanding
this with respect to $\kappa$ we finally have:
$\bar{r}_+=r_+-{\kappa\beta_H m(r_+)/(2r_+)}$,
where the quantities $r_+~,~\beta_H$ are classical  
quantities calculated
for mass $M$ and charge $Q$. From this it immediately follows that
\begin{equation}
\bar{r}^2_+=r^2_+-\kappa\beta_H m(r_+) . 
\label{5.21}
\end{equation}
This identity can be interpreted as  the deformation of the 'horizon  
area' because of the quantum corrections.

\bigskip

\section{Quantum corrections to black-hole thermodynamics}
\setcounter{equation}0

Our approach to the one-loop thermodynamics described by the action  
$W$
(\ref{5.1}) is essentially the same as in the tree-level  
approximation considered in Section 3.
We fix $r_B$, the temperature $T=(2\pi\beta )^{-1}$ on the  
boundary $x=L$  of the system and the black hole topology of the  
space-time geometry,
 and define the off-shell entropy and  
energy by the relations
\begin{equation}
S=(\beta \partial_\beta-1)W,~~~E={1 \over 2\pi} \partial_\beta W
\label{6.5}
\end{equation}
Then, taking the Euclidean static metric
in the form (\ref{1}) with arbitrary functions $g(x), \lambda (x)$  
satisfying the above
conditions ($g(x)$ has simple zero at $x=x_+$), we find the  
equilibrium state of the
system described by the extremum of the functional $W[g(x),r(x), \\  
\lambda (x)]$:
\begin{equation}
\delta W\equiv\delta_r W+\delta_g W +\delta_\lambda W=0 .
\label{6.6}
\end{equation}
For our choice of the action for the quantum field  the one-loop part  
$\Gamma$ does not depend on the dilaton field
$r(x)$. Therefore, a variation of $W$ with respect to $r(x)$ is  
exactly the same as
for the classical action $W_{cl}$, $\delta_r W=\delta_r W_{cl}$ (see  
Eq.(\ref{9})), where now $r(x_+)=\bar{r}_+$ is the quantum value of  
the dilaton field
on the horizon.
This means that the extremum configuration satisfies
the condition
\begin{equation}
{2 \over g'(x_+)} \equiv \beta_H=\bar{\beta} ,
\label{6.7}
\end{equation}
i.e. the extremum as in the classical case is attained on the regular  
manifold without
conical  
singularity\footnotemark\addtocounter{footnote}{0}\footnotetext{Note,  
that in principle the one-loop effective
action $\Gamma$ can be a functional of both $g(x)$ and $r(x)$ leading  
to a more
complicated equation than (\ref{6.7}). In consequence, the extremum
configuration can be singular ($\beta \neq \beta_H$). We do not  
consider this possibility
here.}. 
The extremum of functional $W$ describes the quantum-corrected black  
hole 
configuration the perturbative form of which we found in Section 5  
(Eqs.(\ref{5.9}),
(\ref{5.9a}), (\ref{5.13})).

 In variation with respect to metric, $\delta_g W$, the terms  
depending on $\delta g'(x_+)$
and $\delta g'(L)$ are absent in the same manner as in the classical  
case (see Eq.(\ref{10})). Thus,
for the one-loop effective action $W$ we also have a well-defined  
variational procedure when
the contribution of variations of the normal derivatives of metric at  
the external boundary ($x=L$)
and at the tip of the cone ($x=x_+$) are compensated by the  
corresponding boundary terms.

Calculating the off-shell quantities (\ref{6.5}) it is convenient to  
write metric
in the Schwarz- \\ schild like form:
\begin{equation}
ds^2=g(x)d\tau^2+g^{-1}(x)dx^2 ,
\label{6.1}
\end{equation}
where $0 \leq \tau \leq 2\pi\bar{\beta}$. This always can be done  
using the residual
gauge freedom in Eq.(\ref{1}) allowing choose the coordinate system
where
$\lambda (x)=0$. 
This change of coordinates $x\rightarrow \int^x e^{-\lambda (x)}dx$
must be accompanied by the corresponding change of the integration limits
($x_+,~L$). On-shell they become dependent on $r_B$ and $\beta_H$.
However, for  calculation of  coordinate-invariant off-shell
quantities (like effective action) the using of (\ref{6.1}) instead of general form
(\ref{1})
 is just a
convinient choice of the coordinate system.
The corresponding Polyakov-Liouville action $\Gamma$ reads
\begin{equation}
\Gamma [g]={1 \over 24}\int_{x_+}^{L}({2 \over \bar{\beta} g}  
-{\bar{\beta} \over 2}{g'^2 \over g})
dx+{1 \over 12} (\alpha+{(1-\alpha^2) \over 2\alpha} ) \psi (x_+)
-{\bar{\beta} \over 8}g'(L)+\Gamma_0 ,
\label{6.3}
\end{equation}
where $\alpha=\beta/ \beta_H$;~~$\beta_H=2 [g'(x_+)]^{-1}$, and   
$\psi (x)$ is defined by  Eq.(\ref{IIII}). It should be noted
that (\ref{6.3}) is divergent at the  lower limit.  Taking the  
regularization $x^+
\rightarrow x^++\epsilon$
we have for the divergent part of (\ref{6.3})
\begin{equation}
\Gamma_{div}=\ln \epsilon {(1-\alpha )^2 \over 24 \alpha^2} .
\label{6.4}
\end{equation}
This is the physical divergence due to the infinity of the  
renormalized  $T_{\mu\nu}$
(\ref{3.24}), (\ref{3.25}) at the tip of the cone (for $\bar{\beta}  
\neq \beta_H$).
Note that $\Gamma_{div}$ is proportional to $(1-\alpha)^2$ . Hence  
the divergence does not affect
 physical quantities calculated at the  Hawking temperature  
$(\bar{\beta}=\beta_H )$.
In principle, one can regularize this divergence by subtracting in  
(\ref{6.3}) the Polyakov action
calculated for the Rindler space with metric function
$g_R(x)={2 \over \beta_H}(x-x_+)$. But we do not do this here.

Eq.(\ref{6.7}) allows us to  calculate  the energy  $E$ for the  
equilibrium state
(for $\bar{\beta}=\beta_H$)
$$
E=E_{cl}+E_q,
$$
where the classical part $E_{cl}$ takes the form (\ref{3.5}) while  
the quantum part
reads
\begin{equation}
E_q={1 \over 2\pi g^{1/2}(L)} \partial_{\bar\beta} \Gamma  
|_{\bar{\beta}=\beta_H}=
{1 \over 96\pi}\int_{x_+}^{L}{1 \over g}({4 \over  
\beta^2_H}-g'^2(x))dx
-{1 \over 16\pi g^{1/2}(L)}g'(L) .
\label{6.8}
\end{equation}
For the  quantum-corrected metric obtained in the previous Section  
$g'(L)$
vanishes in the limit $L\rightarrow \infty$. Therefore, we will 
neglect such a term below.

Analogously, we have for the entropy in the equilibrium state
\begin{equation}
S={\pi \bar{r}^2_+ \over G}+S_q,
\label{6.9}
\end{equation}
where
\begin{eqnarray}
S_q&=&(\beta \partial_\beta -1) \Gamma |_{\beta=\beta_H}=-{1 \over  
12} \psi (x_+)
\nonumber \\
&=&{1 \over 12} \int_{x_+}^{L}{dx \over g(x)}({2 \over \beta_H}-g'(x)  
) +{1\over 6}\ln {\beta_H g^{1/2}(L)\over z_0}+c(z_0)~~.
\label{6.10}
\end{eqnarray}
In (\ref{6.9}), (\ref{6.10}) $\bar{r}_+$ and $\beta_H$ are quantum  
position of the
horizon and quantum inverse Hawking temperature respectively and  
$g(x)$ is the metric
of the quantum black hole.
Note that both $E_q$ and $S_q$ are free of  divergences at the lower  
limit.
For a metric written in the conformally flat form  
$g_{\mu\nu}=e^{2\sigma}\delta_{\mu\nu}$,
we have $\psi (x)=-2\sigma (x)$ and the entropy (\ref{6.10})  
coincides with that previously obtained in
\cite{Myers}, \cite{6}, \cite{FS}.

For the energy functional we have:
\begin{equation}
E= {1 \over 2G g^{1/2}(L)}\int_{x_+}^{L}(G^0_0+T^0_0)dx+{1 \over  
12\pi\beta_H g^{1/2}(L)}
+E_{surf} ,
\label{6.11}
\end{equation}
where the surface term $E_{surf}$ is the same as in (\ref{14'}).  

Remembering that the temperature $T=(2\pi \beta_H g^{1/2}(L))^{-1}$  
is fixed on the external boundary we obtain that
when the equations of motion (\ref{5.2}) hold $E$ reduces to
\begin{equation}
E=E_{surf}+{T \over 6} ,
\label{6.12}
\end{equation}
or in  invariant form:
\begin{eqnarray}
E={T \over G}\int_{\partial M}^{}rn^\alpha
\partial_\alpha r  +{T\over 6}=-{1 \over G} \left( r g^{1/2}   
\right)_{r=L}+{T\over 6} .
\label{6.13}
\end{eqnarray}
Note that both the terms in (\ref{6.13}) are defined on the external  
boundary $r=L$.

Subtracting now the energy of the background $g_0$ we obtain:
\begin{equation}
E[g]-E[g_0]={L\over G}\left(r(g^{1/2}_0-g^{1/2})\right)_{r=L}+{1\over  
6}(T-T_0)~~,
\label{V}
\end{equation}
where $T_0=(2\pi \beta^0_H g^{1/2}_0(L))^{-1}$ is the temperature of  
the background metric.
The temperature $T$ which enters Eq. (\ref{6.12}) and (\ref{6.13})
is measured at the external boundary. Nevertheless  the terms which  
contain it originated from the horizon when one integrates by parts  
in passage from Eq.(\ref{6.8}) to
Eq.(\ref{6.11}). Thus, ${T\over 6}$ in (\ref{6.12}), (\ref{6.13}) is  
an consequence
of the black hole topology. In the non-black hole case (hot space)  
this term is absent.
Taking $T_0=T$, the second contribution in (\ref{V}) due to
differences of temperatures vanishes and we get the classical  
expression (\ref{3.18}) for the energy. But now $g$ and $g_0$ are the  
corresponding  quantum corrected metrics.

The above expressions for the energy and entropy were given for the  
static
metric in the form (\ref{6.1}). The quantum-corrected metric found
in Section 5 takes this form by means of the coordinate  
transformation
$r\rightarrow x(r)~,~~\partial_r x=e^{\Phi (r)}$, and identification
$g(x)=fe^{2\Phi}$. 
Since $\Phi (L)=0$, near the boundary $r=L$ we have
$x \approx r$ and $g(L)=f(L)$.

\bigskip

{\bf A. Mass of the quantum-corrected black hole} \\
 The 
 quantum-corrected solution (\ref{5.13}), (\ref{5.14}) found in the  
previous Section  behaves for
 large size $L$ of the box as follows:
\begin{equation}
g(L)\approx 1-{2MG\over L}-{2\kappa\over \beta^2_H}
-{4MG\kappa\over L \beta^2_H}\ln {L\over l}+O({1\over L})~~.
\label{6.14}
\end{equation}
We  see that in the limit $L\rightarrow \infty$ the metric function  
on the boundary
of the box $g(L)$ goes to the constant value $g(L)\rightarrow  
g_0=1-{2\kappa\over\beta^2_H}$
rather than to $1$.
Introducing the Planck temperature $T_{PL}=(2\pi r_{cr})^{-1}$ this  
can be rewritten as $g_0=1-
({T\over T_{PL}})^2$. We see that the modification of the asymptotic  
behavior of $g$ and of the background is essentially due to  
temperature effects. Indeed, if we would take background
 $g_0=1$ as in classics and apply (\ref{V}) for the metric  
(\ref{6.14}) we would obtain 
for the energy the divergent term $E_{th}={\pi\over 6}T^2L$ which is  
the energy of the hot gas surrounding the black hole.

We can interpret  this as follows.
The system under consideration represents  a rather complicated  
interaction
of two objects: black hole and hot gas. Far from the horizon the  
effect
of the gas is more important,  while near the horizon the hole  
dominates.
Therefore,  extensive characteristics (like energy or entropy) of the
system presumably contain different contributions due to these two  
subsystems.
The contribution of the hot gas can be identified and  eliminated  
through its dependence on
the size of the system $L$. On the other hand,
the contribution of the hole itself does not depend on $L$.

It is remarkable fact that the contribution of the hot gas can be  
subtracted and the contribution
of the hole itself  extracted by an appropriate choice of the  
reference
metric\footnote{This
has been demonstrated for the string-inspired 2D model in  
\cite{SS1}.}
in the expression (\ref{V}). Indeed, 
let choose $g_0=1-{2\kappa \beta^{-2}_H}$ for the reference metric.
Then we get for the energy\footnote{ Applying formula (\ref{V}) to  
calculate the energy $E$
we must take into account two different regimes: the perturbative  
expansion in $\kappa$ and the
limit $L\rightarrow \infty$. Therefore, our steps are the following:
we first expand the expression (\ref{6.13}) with respect to $\kappa$
for $L$ fixed and then take the limit $L\rightarrow \infty$.}:
\begin{equation}
E=M+{\kappa M\over \beta^2_H}\eta ,
\label{6.16}
\end{equation}
where $\eta=1+2\ln ({L/l})$.
As we can see the part $E_{th}$ disappeared in (\ref{6.16}), however
the logarithmically divergent term is still present. We think that  
this
divergence is due to the infra-red behavior typical of massless  
fields.
One might expect that it is absent when  massive matter is  
considered.
Therefore, we will keep the size of the box $L$  regularizing this
infra-red behavior to be finite though large enough  with respect to  
the
characteristic size of the hole, $L>>r_+$.
We can rewrite (\ref{6.16}) in the form:
\begin{equation}
E=M(1+{1\over 2}({T\over T_{PL}})^2 \eta) .
\label{6.17}
\end{equation}
Comparing expressions (\ref{6.16}) and (\ref{6.17}) we can conclude  
that 
 the ( first-order (in $G$)) quantum  corrections are identical to  
the  temperature corrections to the mass of the hole.

\bigskip

{\bf B. Entropy of the quantum-corrected black hole} \\
Substituting the classical metric function  
$g_{cl}(r)={(r-r_+)(r-r_-)r^{-2}}$ into the
expression for $S_q$ we find  that
\begin{equation}
S_q={\pi\over 3}T_H(L-r_+)
-{1\over 12}({r_-\over r_+})^2\ln ({L-r_-\over  
r_+ -r_-})
+{1\over 12}\ln ({L-r_+\over r_+ -r_-})+{1\over 6}  
\ln {r_+ \over z_0}~~,
\label{6.18}
\end{equation}
where $z_0$ is the proper cone generator length appearing in Eq.(\ref{3.18}).
Again, as for the calculation of the energy, we observe that $S_q$ is  
divergent in the
limit $L\rightarrow \infty$. The first, linearly divergent, term on  
the r.h.s. of (\ref{6.18})
coincides with the entropy of the 2D hot gas contained in the box  
with size
$(L-r_+)$ and temperature $T_H$, $S_{th}={\pi\over 3}T_H  
(L-r_+)$.
We may subtract the hot gas contribution $S_{th}$ from the expression  
of the
entropy since we are interested in the entropy of the hole itself.

In (\ref{6.9}) the first term is defined with respect to the  
quantum-corrected
radius of the horizon, $\bar{r}_+$.
Near the outer horizon $r=r_+$ the quantum-corrected metric
(\ref{5.13})-(\ref{5.14}) reads as $f(r)={(r-\bar{r}_+)(r-\bar{r}_-)  
r^{-2}}$,  
where $\bar{r}_\pm=r_\pm \pm \kappa r^q_\pm$,
and $r_\pm$ are classical values. 
Therefore, in $S_q$ (which is really proportional to $\hbar$) we may
take the quantum-corrected values $\bar{r}_\pm$ instead of the  
classical  
one.
Then, taking the limit $L\rightarrow \infty$, we derive the complete  
quantum entropy of the hole in terms of the quantum-corrected
horizon values $\bar{r}_\pm$:
\begin{equation}
S={\pi \bar{r}^2_+\over G}
+{1\over 12}(1-({\bar{r}_-\over \bar{r}_+})^2)\ln {L\over  
(\bar{r}_+-\bar{r}_-)}+
{1\over 6}\ln {\bar{r}_+\over z_0}~~.
\label{6.19}
\end{equation}
This illustrates the modification of the classical  {\it "entropy -  
horizon
area"} relation at the quantum level\footnote{One can expect that the  
geometry drastically changes
near the inner horizon $r_-$ due to quantum corrections as was  
previously indicated
in \cite{Trivedi}. As a result, the inner horizon area probably  
becomes a non-analytical
function of the quantum perturbation parameter $\kappa$. Therefore,  
$\bar{r}_-$ in the
expression (\ref{6.19}) is not real inner horizon radius but as it is  
seen from the form of the metric in the region $r \geq r_+$ .}.

A few regimes are of special interest. The first one is the extremal  
limit, $\bar{r}_+ \rightarrow \bar{r}_-$. Note that the correction to  
the mass  
(\ref{6.16}), (\ref{6.17}) vanishes 
then. On the other hand, $S_q$ (\ref{6.18}) has the well-defined  
limit:
\begin{equation}
S^{ext}_q={1\over 6} \ln {\bar{r}_+\over z_0}={1\over 12} \ln  
({A_+\over  
\pi z^2_0})
\label{6.20}
\end{equation}
giving the logarithmic correction to the entropy.
In the other regime we take $\bar{r}_-=0$ (uncharged hole) and get  
for the  
entropy:
\begin{equation}
S={A_+\over 4G}+{1\over 24}\ln {A_+\over \pi z^2_0} ,
\label{6.21}
\end{equation}
where $A_+=\pi \bar{r}^2_+$ is the area of the horizon and we omitted  
a  term $\sim \ln {L/z_0}$. This result is similar to that  obtained in  
\cite{F1} for the four-dimensional Schwarzschild black hole.

It is not quite clear in which phenomena involving black holes the  
logarithmic 
corrections to the entropy (\ref{6.18})-(\ref{6.21}) might be  
important. We may speculate that they play some role in the final  
stage of black hole evaporation. However, this
problem needs further investigation.

\bigskip

\section{Conclusion}
\setcounter{equation}0

In concluding, several remarks are in order. The entropy of a black  
hole in classical theory is determined by data on the horizon surface  
$\Sigma$. In  four-dimensional Einstein gravity it is
just the area of $\Sigma$. In an $R^2$-theory of gravity the entropy  
is given by an integral over $\Sigma$ of the curvature tensors  
projected onto the subspace normal to $\Sigma$ \cite{JKM}, 
\cite{FS}. When the quantum matter contribution is taken into account  
we find, at least in the 2D
case, that the correction to the entropy is also given by some data  
on the horizon, namely by
$\psi |_\Sigma=\psi (x_+)$ (see Eqs.(\ref{6.9})-(\ref{6.10}))  
\cite{Myers}, \cite{FS}.
In fact, the quantum correction contains the contribution of the hot  
gas surrounding
the hole and a
 correction to the entropy of the hole itself. The value of $\psi  
(x_+)$ involves both of them
(see Eqs.(\ref{6.18}), (\ref{6.19})). It  may be unexpected  that   
information on the
hot gas is  encoded in  data at the horizon located far from the  
region where the gas
really contributes. But this becomes less surprising if we recall  
that the function $\psi (x)$
is in fact a non-local object ($\psi=\Box^{-1} R$, see  
eq.(\ref{3.4})) and its value at one
point can, in principle, contain information on the whole space. It  
is not clear whether
this is a general rule, applicable to the four-dimensional case as  
well. In the two-dimensional
model, which is a reduction of the 4D theory, we obtain the one-loop  
entropy of the hole (\ref{6.19})
which is a rather complicated function of the quantum-corrected  
geometry. Probably, there must be an
equivalent derivation of the formula (\ref{6.19}) in terms of the 4D  
geometric (presumably non-local) invariants integrated over $\Sigma$.  

Also it is of interest to make the derivation
of entropy  presented in this paper directly in four dimensions. This  
is a much more
complicated problem. However, some scaling arguments like that given  
in \cite{F1} might be helpful in this project.

\bigskip

\section{Acknowledgments}
The authors thank D.Fursaev for fruitful discussions. The work by
V.Frolov and W.Israel was partly supported  by the Natural Sciences
and Engineering Research Council of Canada.
Research of S.Solodukhin was supported by NATO Fellowship and in part by the Natural
Sciences and Engineering Research Council of Canada.


\begin{thebibliography} \\
\bibitem{1}  J.D.Bekenstein, Lett.Nuov.Cim. {\bf 4}, 737
(1972);
Phys.Rev. {\bf D7},  2333  (1973) ; Phys.Rev. {\bf D9},  3292 (1974).
\bibitem{2} S.W.Hawking, Comm.Math.Phys. {\bf 43}, 199
(1975) .
\bibitem{3} J.M.Bardeen, B.Carter and S.W.Hawking, Comm.Math.Phys.
{\bf 31}, 181 (1973).
\bibitem{4} J.D. Bekenstein, {\it "Do we understand
black
hole entropy?"}, qr-qc/9409015.
\bibitem{5} V.P.Frolov, {\it Black hole entropy and physics at  
Planckian scales}, hep-th/9510156.
\bibitem{6}  T.M.Fiola, J.Preskill, A.Strominger, S.P.Trivedi,  
Phys.Rev. {\bf D50},
3987 (1994).
\bibitem{SS}  S.N. Solodukhin,  Phys.Rev. {\bf D51},  609   (1995).
\bibitem{F1} D.V.Fursaev, Phys.Rev. {\bf D51}, 5352 (1995). 
\bibitem{9} O.Zaslavski, unpublished.
\bibitem{SS1} S.N.Solodukhin, Phys.Rev.{\bf  D53},  (1996), 824;
  hep-th/9506206.
\bibitem{2D}  J.A.Harvey and A.Strominger, {\it "Quantum Aspects of  
Black
Holes"}, Enrico Fermi Institute Preprint (1992), hep-th/9209055;
              S.B.Giddings, {\it "Toy Model for Black Hole  
Evaporation"},
              UCSBTH-92-36, hep-th/9209113;
              R.B.Mann, {\it "Lower dimensional black holes: inside  
and out" },
              WATPHYS-TH-95-02; gr-qc/9501038.
\bibitem{F2} V.P.Frolov, Phys.Rev. {\bf D46}, 5383 (1992).
\bibitem{M}   R.B.Mann, A.Shiekh and L.Tarasov, Nucl.Phys. {\bf  
B341}, 134 (1992).
\bibitem{Polyakov} A.M.Polyakov, Phys.Lett. {\bf B103}, 207 (1981).
\bibitem{FV} V.P.Frolov and G.A.Vilkovisky, Phys.Lett. {\bf B106},  
307 (1981); 
V.P.Frolov and G.A.Vilkovisky, In: Quantum Gravity (Proceedings of  
Second Moscow Quantum Gravity Seminar, Moscow 1981), Eds.Markov M.A.  
and West P.C., Plenum Press, N.Y.-London, 1983.
\bibitem{BD} N.D.Birrell and P.C.W.Davies. Quantum Fields in Curved  
Space. (Cambridge Univ.Press, Cambridge, 1982).
\bibitem{11} L.Susskind, {\it Some Speculations About Black Hole  
Entropy in String Theory}, RU-93-44, hep-th/9309145.
\bibitem{Teitelboim}  S. Carlip and C. Teitelboim, {\it The off-shell
black
hole},
gr-qc/9312002;
             C. Teitelboim, {\it Topological roots of black hole  
entropy},
             preprint, April 1994; M. Ba{\~n}ados, C. Teitelboim and  
J.Zanelli, Phys. Rev. Lett.
             {\bf 72}, 957 (1994).
\bibitem{GH} G.W.Gibbons, S.W.Hawking, Phys.Rev. {\bf D15}, 2752  
(1977).
\bibitem{HH} S.W. Hawking, G. T. Horowitz, {\it The gravitational  
Hamiltonian, action, entropy and surface terms},
DAMTP-R-94-52,  gr-qc/9501014.
\bibitem{hawking} S.W.Hawking in {\it General Relativity}, edited by  
S.W.Hawking
             and W.Israel (Cambridge University Press, Cambridge,  
1979).
\bibitem{FS} D.V.Fursaev, S.N.Solodukhin,   Phys.Rev. {\bf D52}, 2133  
(1995).
\bibitem{Y}  J.W.York, Jr., Phys.Rev. {\bf D33}, 2092 (1986);
B.F.Whiting, J.W.York, Jr., Phys.Rev.Lett. {\bf 61}, 1336 (1988);
H.W.Braden, J.D.Brown, B.F.Whiting and J.W.York, Jr., Phys.Rev. {\bf  
D42}, 3376
(1990).
\bibitem{Kazakov-Solodukhin} D.I.Kazakov, S.N.Solodukhin, Nucl.Phys.  
{\bf B429}, 153 (1994).
\bibitem{EX} S.W. Hawking, G. T. Horowitz and S. F. Ross, Phys.Rev.  
{\bf D51}, 4302 (1995); 
 G.W. Gibbons, R.E. Kallosh, Phys.Rev.{\bf D51}, 2839 (1995);
 C. Teitelboim,  Phys.Rev.{\bf D51}, 4315 (1995).
\bibitem{D}  J.S. Dowker, Class.Quant.Grav. {\bf 11},L7 (1994).
\bibitem{Alvarez} O.Alvarez, Nucl.Phys. {\bf B216}, 125 (1983).
\bibitem{FZ}  V.P.Frolov and A.I.Zelnikov. In: Quantum  
Gravity:Proceedings of the Fourth Seminar on Quantum Gravity, May  
25-29, 1987, Moscow, (Eds.M.A.Markov, V.A.Berezin, and V.P.Frolov,   
World Scientific, Singapore, 1988), p.568.  
\bibitem{W} C. Holzhey, F. Larsen and F. Wilczek, Nucl.Phys. {\bf  
B424}, 443 (1994). 
\bibitem{Reuter} A.H.Chamseddine, M.Reuter, Nucl.Phys. {\bf B317},  
757 (1988).
\bibitem{D1} J.S. Dowker, Phys.Rev.{\bf D50}, 6369 (1994);  
hep-th/9406144.
\bibitem{X} B.Birnir, S.B.Giddings, J.A.Harvey and A.Strominger,
Phys.Rev. {\bf D46}, 638 (1992);
S.W.Hawking, Phys.Rev.Lett., {\bf 69}, 406 (1992);
T.Banks, A.Dabholkar, M.Douglas and M.O.'Loughlin, Phys.Rev. {\bf  
D45}, 3607
(1992).
\bibitem{Lowe} D.Lowe, Phys.Rev. {\bf D47}, 2446 (1993).
\bibitem{Div} W.Israel, Int.J.Mod.Phys. {\bf D3}, 71 (1994);
 D. J. Loranz, W. A. Hiscock  and P. R. Anderson, Phys.Rev. {\bf  
D52}, 4554 (1995);
D.Markovic, E.Poisson, Phys.Rev.Lett. {\bf 74}, 1280 (1995). 
\bibitem{Myers}  R.C.Myers, Phys.Rev. {\bf D50}, 6412 (1994).
\bibitem{Trivedi} S.P.Trivedi, Phys.Rev. {\bf D47}, 4233 (1993).
\bibitem{JKM} T.Jacobson, G.Kang and R.C.Myers, Phys.Rev. {\bf D49},  
6587 (1994).
\end{thebibliography}
\end{document}